\documentclass{jfm}
\usepackage{graphicx}
\usepackage{newtxtext}
\usepackage{newtxmath}
\usepackage{natbib}
\usepackage{hyperref}
\usepackage{bm}
\usepackage{caption}
\usepackage[normalem]{ulem}
\usepackage[dvipsnames]{xcolor}
\hypersetup{
    colorlinks = true,
    urlcolor   = blue,
    citecolor  = CadetBlue,
}

\title{Influence of plume activity on thermal convection in a rectangular cell}

\author{Ambrish Pandey\aff{1,2},
        \corresp{\email{ambrish.pandey@ph.iitr.ac.in}}
        J\"org Schumacher\aff{3,4},
        Matteo Parsani\aff{5},
        \and Katepalli R. Sreenivasan\aff{2,4,6}
        }

\affiliation{\aff{1} Department of Physics, Indian Institute of Technology Roorkee, Roorkee 247667, Uttarakhand, India
\aff{2} Center for Astrophysics and Space Science, New York University Abu Dhabi, Abu Dhabi 129188, United Arab Emirates \\
\aff{3} Institute of Thermodynamics and Fluid Mechanics, Technische Universität Ilmenau, D-98684 Ilmenau, Germany
\aff{4} Tandon School of Engineering, New York University, New York, NY 11201, USA
\aff{5} Computer, Electrical and Mathematical Sciences and Engineering Division, King Abdullah University of Science and Technology (KAUST), 23955-6900, Thuwal, KSA
\aff{6} Department of Physics and Courant Institute of Mathematical Sciences, New York University, New York, NY 11201, USA
}

\begin{document}
\maketitle


\begin{abstract}
We present three-dimensional direct numerical simulations of turbulent Rayleigh-B\'enard convection in a closed rectangular box whose width $L_y$ and length $L_x$ are 0.8 and 2.4 times the height $H$, respectively. The Rayleigh number $Ra$ varies from $10^5$ to $10^{10}$, and the Prandtl number is unity. The advantages of the present configuration are: (a) A relatively stable unidirectional large-scale circulation, consisting of two counter-rotating rolls, fills the cell and fixes the thermal plume ejection- and shear-dominated regions, in contrast to those in closed cylindrical cells. (b) The regions of plume ejection are essentially independent of the sidewalls so that their autonomous existence can be studied. This is because there is some space, or "fetch",  for the velocity and thermal boundary layers to develop along the length. (c) This geometry allows one to study the influence of locally thin and thick boundary layers (which follow larger or smaller plume activity) on the scaling of convection properties. In regions of larger plume activity (defined by an incessant movement of plumes), the temperature fluctuation as well as the normalised thermal and viscous dissipation rates decay more slowly with $Ra$ than in regions of lower activity. Both viscous and thermal boundary layers thin down rapidly with increasing distance from the plume ejection region. The local thicknesses of {both} boundary layers decline more rapidly with $Ra$ in the ejection region than in regions of impact and shear, where they are similar to each other. Despite these details, the global heat transport laws are practically the same as those in other configurations of low to moderate aspect ratios.
\end{abstract}


\section{Introduction} 
\label{sec:intro}

An outstanding question in thermal convection is the dynamical interaction of the flow close to the boundaries, leading to the emission of plumes, with the rest of the flow. For large aspect ratios, thermal boundary layers can be identified even at very high Rayleigh numbers, whereas a unidirectional momentum boundary layer does not seem to exist \citep{Samuel:JFM2024, Shevkar:PNAS2025,Samuel:FDR2025}. For small aspect ratios, the emergence of a strong mean wind~\citep{Sreenivasan:PRE2002,Ahlers:RMP2009,Shi:JFM2012,Lohse:RMP2024} maintains a semblance to the momentum boundary layer---although, because of the small ``fetch" and the up- and down-welling regions near the bounding vertical walls, the flow bears little resemblance to the canonical form. Moreover, the direction of the mean wind drifts in cylindrical cells~\citep{Sreenivasan:PRE2002, Xi:PRE2006, Mishra:JFM2011, Schumacher:PRF2016} as well as horizontally-periodic~\citep{Reeuwijk:PRE2008a} domains.

\begin{figure}
\captionsetup{width=1\textwidth}
\centering
\includegraphics[width=0.95\textwidth]{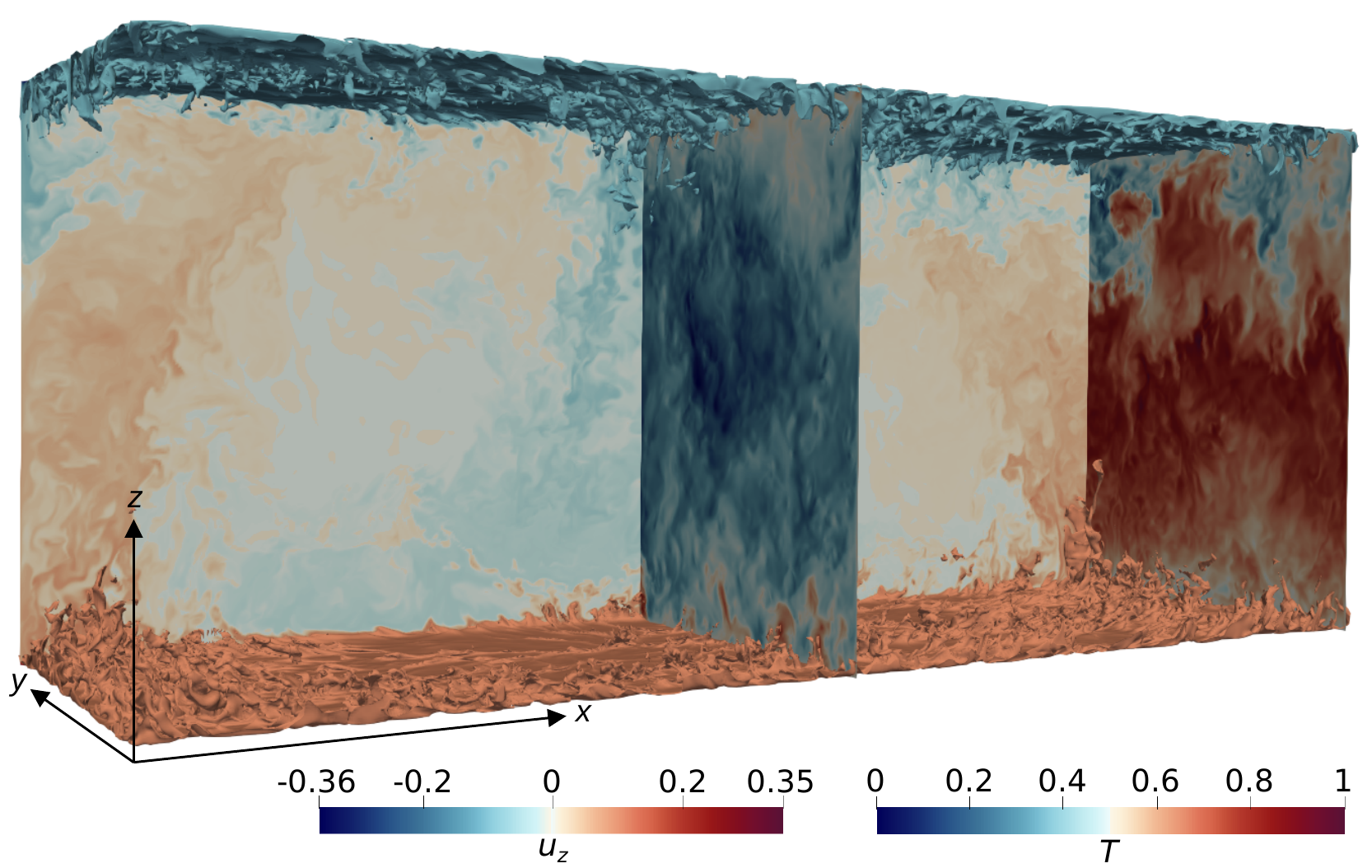} 
\caption{Instantaneous flow realisation for $Ra = 10^{10}$ in the closed rectangular cuboid with the red and blue colours representing, respectively, hot upwelling and cold downwelling structures. Fine thermal structures are revealed by temperature isosurfaces: hot and cold structures correspond to $T = 0.65$ and $T = 0.35$, respectively. Vertical velocity contours, shown in the vicinity of the sidewall at $x \approx L_x$ and on the mid-vertical plane at $x = L_x/2$, as well as temperature contours on the sidewall at $y \approx L_y$ are consistent with the double-roll structure spanning the width of the domain.}
\label{fig:flow_vis}
\end{figure}

\begin{figure}
\captionsetup{width=1\textwidth}
\centering
\includegraphics[width=0.85\textwidth]{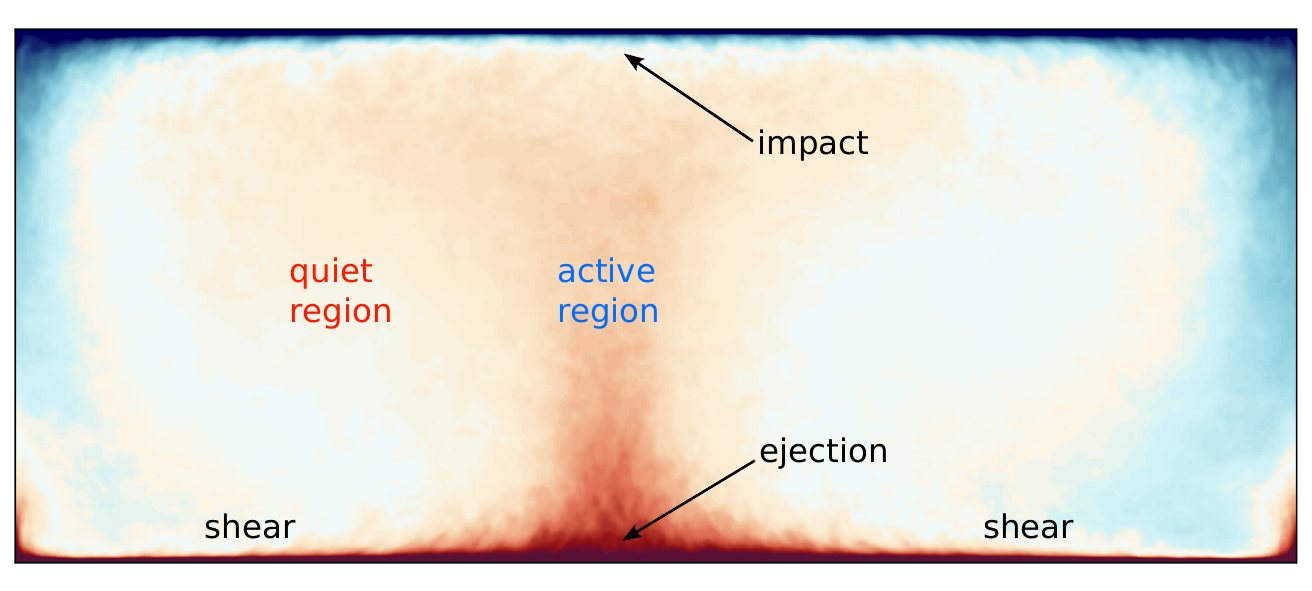}
\caption{An illustration depicting various regions of the flow in the present configuration, using a representative mean temperature field in $y = L_y/2$ plane for $Ra = 10^9$. Ejection, impact and shear regions are observed at the horizontal plates, while active and quiet regions exist deeper in the bulk of the convective layer.}
\label{fig:sketch}
\end{figure}

However, if the flow has a longer fetch in a horizontal direction, the near-wall flow could be closer to canonical wall-bounded flows, at least in the mid-region. If a flow configuration with a well-defined mean flow can be identified, it would be worthwhile understanding its dynamics. To this end, Rayleigh-B\'{e}nard convection (RBC) in a rectangular cuboid of length $L_x=2.4H$, width $L_y=0.8H$, and height $H$ is studied here, in the configuration shown in figure~\ref{fig:flow_vis}. As identified schematically in figure~\ref{fig:sketch}, the flow is characterized by ejection regions in which the thermal plumes emanate, moving essentially upwards into the bulk of the flow; impact regions that receive vertically moving thermal plumes traversing from the opposite plate; and shear regions in which the plumes after impact travel mostly parallel to the plate.

It is already known that the boundary layer properties in the ejection, shear and impact regions are different \citep{He:PRL2019}. Two-dimensional (2D) models of RBC also reveal that the properties of thermal boundary layer (TBL) in the ejection region differ substantially from those in the shear and impact regions~\citep{Poel:PRL2015}, with temperature profiles in the ejection region exhibiting a logarithmic behaviour, which is absent in the shear and impact regions~\citep{Pandey:JFM2021, He:POF2024}. The ejection and impact regions at the horizontal plates affect flow properties within the boundary layers as well as those deeper into the bulk~\citep{Xu:JFM2024,He:PRL2007,Chong:JFM2016,Huang:JFM2016}.

In this paper, we perform direct numerical simulations (DNS) of 3D convection for $Pr = 1$ and $Ra$ up to $10^{10}$ in a rectangular box of aspect ratios $\Gamma_x = L_x/H = 2.4$ and $\Gamma_y = L_y/H = 0.8$. As the first task, we explore the large-scale structure, and find it to be two horizontally stacked cells over the entire domain, with fixed orientations for the entire duration of simulations (see figure~\ref{fig:flow_vis}). This flow morphology results in nearly well defined and spatially fixed ejection, shear, and impact regions on the bottom and top plates of the cell. Further, deeper in the bulk, the region near the mid length experiences continuous movement of thermal plumes and is thus an active region, whereas regions about a quarter length from the walls are relatively quiet; see figure~\ref{fig:sketch}. As a second task, we explore the characteristics of the flow in the bulk and the boundary layer corresponding to the various conditions stated above, especially the scaling characteristics of velocity and thermal fluctuations. One of our findings in this respect is that, despite these differences, the global transports of heat and momentum are practically identical, within error bars, to the one in horizontally unconstrained layer configurations \citep{Samuel:JFM2024}. 

The outline of the manuscript is as follows. \S\ref{sec:sim} is an account of the numerical method---the account is brief because it is by now standard---and a summary of simulation parameters. \S\ref{sec:flow_glob} considers global transport in regions of high and low plume activity, while \S\ref{sec:BL} considers boundary layers. The paper concludes with summary remarks in \S\ref{sec:concl}. A major finding of the present work is that the scaling properties depend strongly on whether or not one considers regions of high or low plume activity, with the temperature fluctuations and the normalised dissipation rates decaying more slowly with the Rayleigh number in regions of higher activity.

\section{Numerical method} 
\label{sec:sim}

The following non-dimensionalized governing equations of RBC are integrated in time:
\begin{eqnarray}
\frac{\partial {\bm u}}{\partial t} + {\bm u} \cdot \nabla {\bm u} & = & - \nabla p + T \hat{z} + \sqrt{\frac{Pr}{Ra}}   \nabla^2 {\bm u}, \label{eq:u} \\ 
\frac{\partial T}{\partial t} + {\bm u} \cdot \nabla T & = &  \frac{1}{\sqrt{Ra Pr}} \nabla^2 T, \label{eq:T} \\ 
\nabla \cdot {\bm u} & = & 0. \label{eq:m}
\end{eqnarray}
Here, ${\bm u} = (u_x, u_y, u_z)$, $T$, and $p$ are the velocity, temperature, and pressure fields, respectively. These equations are non-dimensionalized using $H$, $\Delta T$, the free-fall velocity $u_f = \sqrt{\alpha g \Delta T H}$ and the free-fall time $t_f = H/u_f$ as the length, temperature, velocity, and time scales, respectively. The Rayleigh and Prandtl numbers are defined as $Ra = \alpha g \Delta T H^3/(\nu \kappa)$ and $Pr = \nu/\kappa$; $\Delta T$ is the temperature difference between the bottom and top plates; $\alpha, \nu, \kappa$ are respectively the isobaric thermal expansion coefficient, kinematic viscosity, and thermal diffusivity; $g$ is the acceleration due to gravity. 

\begin{table}
\captionsetup{width=1\textwidth}
  \begin{center}
\def~{\hphantom{0}}
  \begin{tabular}{lcccccccccccc}
 $Ra$ & $N_e \times N^3$ & $Nu$ & $Nu_{\varepsilon_u}$ & $Nu_{\varepsilon_T}$ &  $Nu_W$ & $Re$ & $t_{sim} \, (t_f)$ \\
$3\times 10^{5}$ & $39936\times 3^3$ 	 & 	  5.78 $\pm$    0.4 &   5.77 $\pm$    0.4 &   5.78 $\pm$    0.3 &   5.78 $\pm$    0.3 &     96 $\pm$     25 &   1491 \\ 
$6\times 10^{5}$ & $39936\times 3^3$ 	 & 	  7.17 $\pm$    0.5 &   7.16 $\pm$    0.4 &   7.16 $\pm$    0.3 &   7.17 $\pm$    0.4 &    135 $\pm$     38 &   1470 \\ 
        $10^{6}$ & $39936\times 5^3$ 	 & 	  8.44 $\pm$    0.6 &   8.44 $\pm$    0.5 &   8.44 $\pm$    0.4 &   8.44 $\pm$    0.4 &    172 $\pm$     50 &   1273 \\ 
$3\times 10^{6}$ & $39936\times 7^3$ 	 & 	 11.72 $\pm$    0.9 &  11.73 $\pm$    0.7 &  11.73 $\pm$    0.5 &  11.72 $\pm$    0.5 &    299 $\pm$     85 &   1054 \\ 
        $10^{7}$ & $39936\times 7^3$ 	 & 	 16.53 $\pm$    1.2 &  16.53 $\pm$    0.8 &  16.52 $\pm$    0.6 &  16.53 $\pm$    0.6 &    548 $\pm$    141 &   1073 \\ 
$3\times 10^{7}$ & $145920\times 7^3$ 	 & 	 22.39 $\pm$    1.5 &  22.39 $\pm$    1.1 &  22.40 $\pm$    0.7 &  22.40 $\pm$    0.7 &    949 $\pm$    210 &    345 \\ 
        $10^{8}$ & $145920\times 9^3$ 	 & 	 31.39 $\pm$    2.0 &  31.37 $\pm$    1.3 &  31.38 $\pm$    0.8 &  31.38 $\pm$    0.8 &   1711 $\pm$    377 &    395 \\ 
$3\times 10^{8}$ & $301056\times 7^3$ 	 & 	 42.99 $\pm$    2.6 &  42.98 $\pm$    1.9 &  42.99 $\pm$    0.9 &  42.98 $\pm$    1.0 &   2923 $\pm$    635 &    322 \\ 
        $10^{9}$ & $301056\times 11^3$ 	 & 	 61.56 $\pm$    3.8 &  61.20 $\pm$    2.6 &  61.57 $\pm$    1.1 &  61.57 $\pm$    1.3 &   5261 $\pm$   1189 &    160 \\ 
       $10^{10}$ & $6030704\times 7^3$ 	 & 	 125.3 $\pm$      7 &  124.9 $\pm$      7 &  125.3 $\pm$      2 &  125.3 $\pm$      2 &  16260 $\pm$   3458 &    126 \\  
  \end{tabular}
  \caption
  {Parameters of DNS conducted in a domain of dimensions $L_x:L_y:H = 2.4:0.8:1$ for $Pr = 1$. Total number of mesh cells $N_e \times N^3$ in the flow domain; Nusselt numbers $Nu$, $Nu_{\varepsilon_u}$, $Nu_{\varepsilon_T}$, and $Nu_W$ computed using equations~\eqref{eq:Nu}, \eqref{eq:Nu_epsv}, \eqref{eq:Nu_epst}, and \eqref{eq:Nu_wall}, respectively; Reynolds number $Re$ computed using equation~\eqref{eq:Re}; and the total integration time $t_{sim}$ in the statistically steady state. Nusselt numbers computed using various methods agree to within 1\% for all the simulations, indicating a good spatial and temporal convergence~\citep{Pandey:JFM2025}. Error bars in global quantities represent the standard deviation.} 
  \label{table:sim_detail}
  \end{center}
\end{table}

We conduct DNS of RBC for $Pr = 1$ and $Ra$ up to $10^{10}$ using an open-source solver {\sc Nek5000}~\citep{Fischer:JCP1997, Scheel:NJP2013}, which utilises the spectral element scheme. The rectangular box is decomposed into a finite number of elements $N_e$ and each element is refined further by using a spectral interpolation with Legendre polynomials of order $N$ in all directions. Thus, the flow domain is discretized using $N_e N^3$ mesh cells. Velocity field satisfies the no-slip condition on all the walls. For the temperature field, adiabatic condition is imposed on the sidewalls and isothermal condition on the horizontal plates. Mesh clustering near all the walls is used to capture rapid spatial variation of temperature and velocity fields. The flow is allowed to evolve under the Oberbeck-Boussinesq approximation.

Crucial parameters of DNS are listed in table~\ref{table:sim_detail}. We ensure that the ratio of local vertical grid spacing and the Kolmogorov scale $\eta(z)$ remains lower than 1.8~\citep{Pandey:PD2022}, where $\eta =(\nu^3 / \varepsilon_u)^{1/4}$ is nominally the finest scale in the velocity field and $\varepsilon_u$ is the kinetic energy dissipation rate computed as
\begin{equation}
\varepsilon_u (\bm x) = \frac{\nu}{2} \sum_{l,m=1}^3 \left( \frac{\partial u_l}{\partial x_m} + \frac{\partial u_m}{\partial x_l} \right)^2 \label{eq:eps_u},
\end{equation}
with $l, m \in \lbrace x,y,z \rbrace$. The thermal dissipation rate is computed as
\begin{equation}
\varepsilon_T (\bm x) = \kappa \sum_{j=1}^3 \left(  \frac{\partial T}{\partial x_j} \right)^2 \, . \label{eq:eps_T}
\end{equation}

Global dissipation rates, i.e., the volume- and time-averaged values, are related to global heat transport in RBC~\citep{Howard:ARFM1972}.  
The Nusselt number $Nu$, which is the ratio of the total heat transport to that only by conduction, is computed as 
\begin{equation}
Nu = 1 + \sqrt{Ra Pr} \, \langle u_z T \rangle_{V,t} \label{eq:Nu}
\end{equation}
where $\langle \cdot \rangle_{V,t}$ is the combined volume-time average. Nusselt numbers computed from the dissipation rates are defined as
\begin{eqnarray}
    Nu_{\varepsilon_u} & = & \sqrt{Ra Pr} \, \langle \varepsilon_u \rangle_{V,t} + 1 \, , \label{eq:Nu_epsv} \\
    Nu_{\varepsilon_T} & = & \sqrt{Ra Pr} \, \langle \varepsilon_T \rangle_{V,t} \, . \label{eq:Nu_epst}
\end{eqnarray}
The Nusselt number at the walls is computed from the vertical derivative of temperature:
\begin{equation}
    Nu_W = -  \left. \frac{ \partial \langle T \rangle_{A,t}} {\partial z} \right| _{z=0,H} \, . \label{eq:Nu_wall}
\end{equation}
The heat transport computed using different methods agree to within 1\% (see table~\ref{table:sim_detail}). This concurrence strongly indicates that the resolution is adequate for faithful DNS~\citep{Stevens:JFM2010, Shishkina:NJP2010}. Note that the computational constraints limit the integration time that could be realised for simulations at higher Rayleigh numbers. However, to ensure the adequacy of the duration of simulations, we have checked the convergence of global quantities by computing the difference between the two halves of the dataset. We find that the differences in results remain of the order of 2 percent (in some cases a tenth as small).

\section{Global transport, flow structure and fluctuations in high and low plume activity regions of the bulk flow}
\label{sec:flow_glob}

\subsection{Global transports of heat and momentum}
\label{sec:global}

\begin{figure}
\captionsetup{width=1\textwidth}
\centering
\includegraphics[width=1\textwidth]{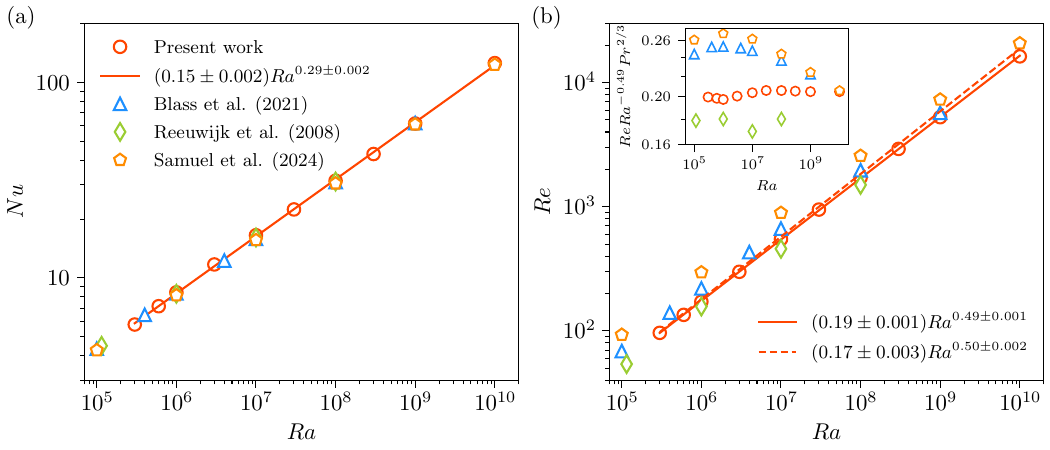} 
\caption{(a) Global heat transport, $Nu$, increases with the thermal driving as $Ra^{0.29}$ in the rectangular box. Present results agree excellently with those for $Pr = 1$ from \citet{Blass:JFM2021a} and \citet{Reeuwijk:PRE2008a} obtained respectively in $\Gamma = 32$ and $\Gamma = 4$ horizontally-periodic square cuboids as well as those for $Pr = 0.7$ from \citet{Samuel:JFM2024}. (b) Reynolds number based on $u_{\rm RMS}$ increases almost as $\sqrt{Ra}$. Dashed line stands for the scaling of the Reynolds number based on $U_{\rm max}$. Inset shows the normalised Reynolds number $Re Ra^{-0.49} Pr^{2/3}$ as a function of $Ra$. The scaling exponents agree well with those observed in literature but the prefactor depends on the shape and size of the convective layer.}
\label{fig:nure}
\end{figure}

The variation of $Nu$ with $Ra$ in the rectangular box is shown in figure~\ref{fig:nure}(a). The best fit to the data in this range of $Ra$ follows $Nu = 0.15 Ra^{0.29}$. We also plot the Nusselt numbers from DNS studies in horizontally-periodic square cuboids of $\Gamma = 32$~\citep{Blass:JFM2021a} and $\Gamma = 4$~\citep{Reeuwijk:PRE2008a}, both for $Pr = 1$, as well as data for $Pr = 0.7$ in a horizontally-periodic cuboid of $\Gamma = 4$~\citep{Samuel:JFM2024}. $\Gamma = L/H$ is the aspect ratio. This power-law exponent is close to $2/7$ obtained in the mixing-zone model~\citep{Castaing:JFM1989} and to that by the turbulent boundary layer model~\citep{Shraiman:PRA1990, Siggia:ARFM1994}.  We conclude that the heat transport is largely insensitive to the aspect ratio (for $\Gamma \geq 0.5$) and the shape of the convection cell, and also to whether the domain is closed or horizontally periodic. The mixing model of \citet{Castaing:JFM1989} is linked to boundary layer only indirectly by being the source of plumes, whereas the Shraiman-Siggia model~\citep{Shraiman:PRA1990} relies strongly on our empirical understanding of the turbulent boundary layer. The two models do not share much physics for a first look, yet they produce the same result on the $Nu-Ra$ relation. A plausible inference to draw from this observation is that the dynamical aspects of boundary layers and the mixing regions are intimately intertwined for these Rayleigh numbers.

We now consider the Reynolds number $Re$, which is commonly estimated using the root mean square (RMS) velocity defined as
\begin{equation}
u_{\rm RMS} = \sqrt{\langle u_x^2 + u_y^2 + u_z^2 \rangle_{V,t}} \, .
\end{equation}
The Reynolds number based on $u_{\rm RMS}$ and $H$,
\begin{equation}
Re = u_{\rm RMS} \sqrt{Ra/Pr} \, , \label{eq:Re}
\end{equation}
is plotted in figure~\ref{fig:nure}(b) as a function of $Ra$. The best fit to the data yields $0.19 Ra^{0.49}$. We also compute the Reynolds number based on $U_{max}$, the maximum horizontal velocity observed close the horizontal plates ( \citet{Niemela:JFM2001, Lam:PRE2002}; see \S~\ref{sec:mean_BL} for a concrete definition), and find good agreement with equation~\eqref{eq:Re}. The best fit to the data shows that the $U_{max}$-based $Re$ follows $0.17 Ra^{0.50}$. A discernible difference between the two Reynolds numbers occurs only at high Rayleigh numbers, for which $U_{max}$ is slightly higher than $u_{\rm RMS}$.

For comparison, data from \citet{Blass:JFM2021a} and \citet{Reeuwijk:PRE2008a} are included in figure~\ref{fig:nure}(b), for which the best fits are $0.29 Ra^{0.48}$ and $0.18 Ra^{0.49}$. Reynolds numbers from \citet{Samuel:JFM2024} are systematically larger---presumably because their configuration is horizontally periodic (with the aspect ratio of 4). We also show the normalised Reynolds number $Re Ra^{-0.49} Pr^{2/3}$ in the inset of figure~\ref{fig:nure}(b). Though the Reynolds numbers in the closed cuboid are smaller than those in the horizontally-periodic cuboids, the data from \citet{Samuel:JFM2024} and \citet{Blass:JFM2021a} are close.  The scaling exponent of $Re$ in different convection domains is about 0.50, but the prefactors are different: the magnitude of the Reynolds number, not its scaling, depends on the shape and size of the container. The $Re-Ra$ scaling in the present work also agrees with those found for moderate-$Pr$ fluids~\citep{Scheel:JFM2012, Wagner:JFM2012, Pandey:POF2016, Pandey:JFM2022}. 

\subsection{Flow organisation}
\label{sec:flow_str}

\begin{figure}
\captionsetup{width=1\textwidth}
\centering
\includegraphics[width=\textwidth]{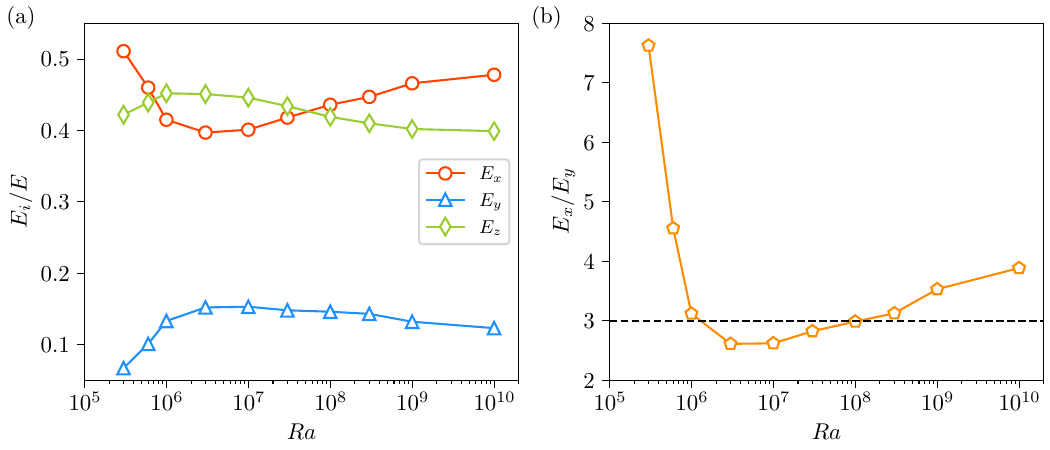}
\caption{(a) Fractions of the kinetic energy corresponding to velocity components as functions of $Ra$ exhibit non-monotonic behavior. The weakest component is $E_y$, whereas $E_x \approx E_z$, which indicates that the flow evolves primarily in the $xz$-plane. (b) Ratio $E_x/E_y$ as a function of $Ra$; after a rapid initial decline, $E_x/E_y$ grows from $2.5$ to 4 for $Ra$ between $10^7$ and $10^{10}$, implying that the flow becomes increasingly horizontally anisotropic. The horizontal line indicates that $E_x$ remains more or less close to $3 E_y$ for a wide range of $Ra$. 
}
\label{fig:ke}
\end{figure}

We compute the global kinetic energy corresponding to each velocity component,  $E_x = \langle u_x^2/2 \rangle_{V,t}, E_y = \langle u_y^2/2 \rangle_{V,t}, E_z = \langle u_z^2/2 \rangle_{V,t}$, and plot in figure~\ref{fig:ke}(a) their ratios with respect to the total energy $E = E_x + E_y + E_z$.  $E_x$ is systematically larger than $E_y$, suggesting that the flow is stronger in the $xz$-plane; this results from our convection domain being longer in the $x$-direction than in $y$. The ratio $E_x/E_y$ plotted in figure~\ref{fig:ke}(b) shows that $E_x \gg E_y$ at the lowest $Ra$ (it is about 30 for $Ra = 10^5$, not shown here), suggesting that the flow is nearly two-dimensional at low Rayleigh numbers, as in extended and weakly forced convection~\citep{Busse:RPP1978, Pandey:NComm2018}. The flow becomes approximately three-dimensional in the $Ra$-range between $10^6$ and $3 \times 10^6$, as can be evidenced by the evolution of $E_x$, $E_y$, and $E_z$ with $Ra$ in this region. The asymptotic ratio $E_x \approx 3 E_y$ is in the same proportion to the ratio of the configurational geometry.

In the following, we restrict comments on the 3D range of Rayleigh numbers, even though we present data for all the Rayleigh numbers considered. 

A snapshot of the flow shown in figure~\ref{fig:flow_vis} reveals two counter-rotating large-scale structures stacked along the $x$-direction. In the region $x \approx L_x/2$, the flow at the bottom plate comprises hot rising plumes for some Rayleigh numbers, and cold falling plumes for others. Thus, in this study, the ejection region at the bottom plate is located at $x \approx L_x/2$ and the impact region at $x \approx 0$ or $ L_x$ depending upon the specific $Ra$. Irrespective of $Ra$, regions $x \approx L_x/4$ and $x \approx 3 L_x/4$ on both bottom and top plates correspond to large shear; here plumes are primarily carried along with the large-scale structures parallel to the horizontal plates. See also figure~\ref{fig:sketch}. The stable two-roll structure establishes a mean flow in the rectangular cuboid having a dominant $x$-component of velocity.

\subsection{Fluctuations in high and low activity bulk regions}
\label{sec:fluct}

Hot or cold plumes are continuously emitted in the region $x \approx L_x/2$ (see figures~\ref{fig:flow_vis} and~\ref{fig:sketch}) and their emission affects the flow through the bulk of the flow. This leads us to explore the position-dependent flow characteristics in the bulk. As the flow evolves primarily in the $xz$-plane, we choose three cubic regions of volume $V_b = (0.4 H)^3$ nested in the bulk, with the regions centered at fixed $y_b = L_y/2, z_b = H/2$, but varying $x_b = \lbrace L_x/4, L_x/2, 3L_x/4 \rbrace$. We compute instantaneous RMS velocity and temperature as well as mean kinetic energy and thermal dissipation rates in the bulk regions and find that all the quantities in quiet regions (at $x_b = L_x/4$ and $x_b = 3L_x/4$) fluctuate about nearly the same mean. However, those in active region at $x_b = L_x/2$ fluctuate about a higher mean. The local flow states at $x_b = L_x/4$ and $x_b = 3L_x/4$ are similar and differ from that at $x_b = L_x/2$. This behavior is true for all Rayleigh numbers, so we combine those data for $x_b = L_x/4$ and $x_b = 3L_x/4$. For reference, we call attention to figure~\ref{fig:sketch} for the locations of the quiet and active regions.

\begin{figure}
\captionsetup{width=1\textwidth}
\centering
\includegraphics[width=1\textwidth]{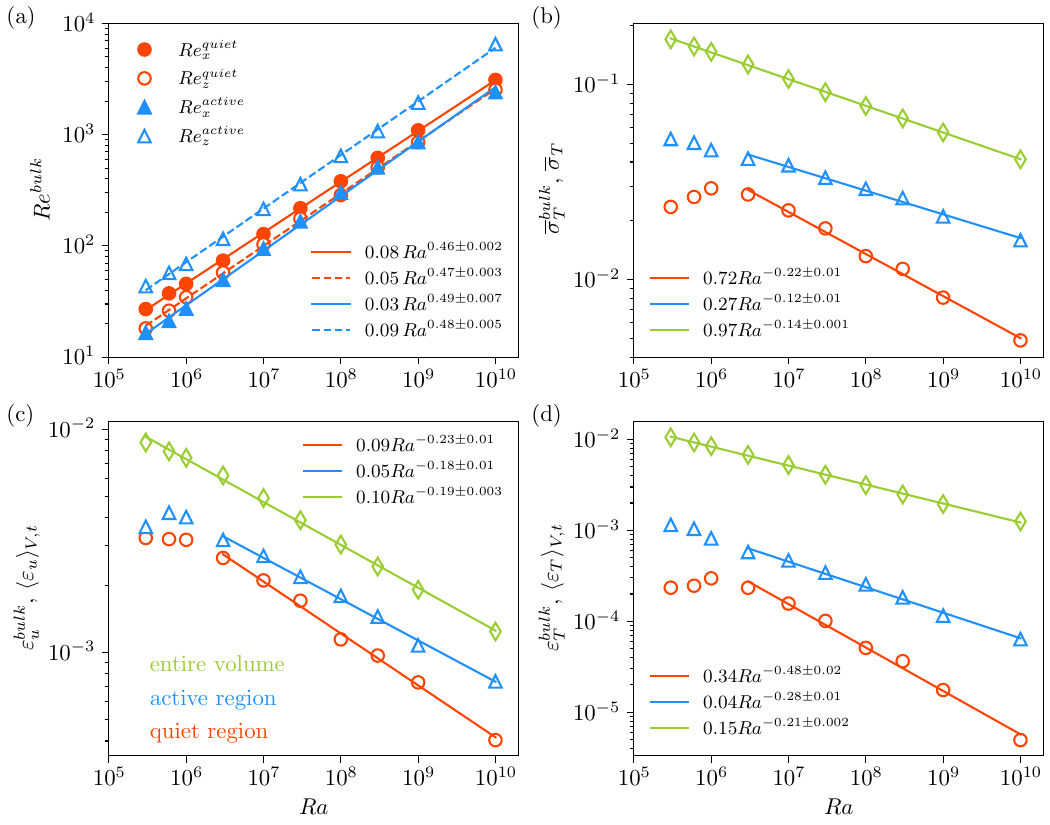}
\caption{Scaling relations in quiet (red) and active (blue) bulk regions. (a) Reynolds numbers based on the $x$-component (filled symbols) and $z$-component (open symbols) of velocity. RMS temperature fluctuation (b), kinetic energy dissipation rate (c), and thermal dissipation rate (d) are larger, and decrease slowly with $Ra$, in the active region than those in the quiet region. Green diamonds in panels (b), (c), (d) are the corresponding quantities in the entire domain.}
\label{fig:bulk_Ra}
\end{figure}
 
In figure~\ref{fig:bulk_Ra}(a), we show the Reynolds numbers in the bulk regions based on the $x$- and $z$-components of velocity, computed as
\begin{equation}
Re_x^{bulk} = 0.4 \, \sqrt{ \langle u_x^2 \rangle_{V_b,t} \, Ra/Pr} \quad  \mbox{and} \quad  Re_z^{bulk} = 0.4 \,\sqrt{ \langle u_z^2 \rangle_{V_b,t} \, Ra/Pr} \, .
\end{equation}
(The factor 0.4 arises because the side length of $0.4 H$ is used to define the Reynolds number.) The relative magnitudes of these Reynolds numbers depend on the region being considered. In quiet regions, $Re_x^{bulk}$ remains larger than $Re_z^{bulk}$, whereas, in the active region, $Re_z^{bulk}$ is significantly larger than $Re_x^{bulk}$. The best fits for $Re_x^{bulk}$ are $Ra^{0.46}$ and $Ra^{0.49}$ in the quiet and active regions, respectively. Similarly, $Re_z^{bulk}$ scales as $Ra^{0.47}$ and $Ra^{0.48}$, respectively, in quiet and active regions. These findings suggest that the scaling of $Re^{bulk}$ is nearly the same in the entire domain, and the increase of $Ra$ affects the velocity fluctuations in all regions of the flow similarly~\citep{Xu:JFM2024}.

We now compute the RMS temperature for the entire domain as
\begin{equation}
\overline{\sigma}_T = \sqrt{\langle T^2 \rangle_{V,t} - \langle T \rangle_{V,t}^2}  \,  \label{eq:sigmaT}
\end{equation}
and show its variation with $Ra$ in figure~\ref{fig:bulk_Ra}(b): $\overline{\sigma}_T$ decreases as $0.97Ra^{-0.137}$, consistent with  \citet{Castaing:JFM1989, Wu:PRA1992,Niemela:Nature2000}, and also with~\cite{Pandey:JFM2025} in a 2D square for $Pr = 1$. This feature seems to be common for all known constrained cases. The reason for decreasing $\overline{\sigma}_T$ with $Ra$ is that the plumes become finer and occupy increasingly smaller volume, which implies that increasingly larger fractions of fluid volume possess temperatures close to the mean and contribute weakly to $\overline{\sigma}_T$. 

The RMS temperature in the bulk regions, $\overline{\sigma}_T^{bulk}$, is computed by using $\langle \rangle_{V_b}$ in place of $\langle \rangle_{V}$ in equation~\eqref{eq:sigmaT}. Figure~\ref{fig:bulk_Ra}(b) shows that it remains larger in the active region, for which the best fit to the data yields $Ra^{-0.12}$, similar to that found by \citet{Samuel:JFM2024}. The best fit to the data in quiet region is $\overline{\sigma}_T \sim Ra^{-0.21}$, decaying substantially faster than in the active region.

If the observed trends for temperature fluctuation in various regions are extrapolated, the RMS will be dominated increasingly by contribution from active regions. This implies that the thermal plumes generated near the horizontal plates are swept away by the dominant double-roll structure more effectively as $Ra$ increases. This is plausible because the strength of the dominant flow structure increases with $Ra$~\citep{Lui:PRE1998, Niemela:EPL2003, Chandra:PRL2013, Pandey:JFM2021} in both 2D and 3D. The faster decrease of $\overline{\sigma}_T^{bulk}$ in the quiet region is also compatible with this picture as the thermal structures arriving in this region become less probable.

The kinetic energy dissipation rates in the entire cell as well as in the bulk regions are shown in figure~\ref{fig:bulk_Ra}(c). The best fit for the former is $\langle \varepsilon_u \rangle_{V,t} = 0.10 Ra^{-0.19}$, consistent with the Nusselt number scaling shown in figure~\ref{fig:nure}(a) (using the exact relation~\eqref{eq:Nu_epsv}). The dissipation in the active region scales the same as that of the global value, $\varepsilon_u^{bulk} = 0.05 Ra^{-0.18}$, though smaller by a factor of about two. 
In the quiet region, $\varepsilon_u^{bulk}$ decays more rapidly with $Ra$, the best fit being $0.09 Ra^{-0.23}$. Taken together, figure~\ref{fig:bulk_Ra}(c) suggests that with increasing $Ra$, $\varepsilon_u^{bulk}$ in the active region slowly approaches the global dissipation. \citet{Li:PRF2021} found, using Lagrangian measurements, that the normalised local dissipation rate in plume-abundant regions scales as $Ra^{1.34}$. Our data yield $Ra^{1.32}$ in the active region and $Ra^{1.27}$ in the quiet region. 

We show the global and bulk thermal dissipation rates in figure~\ref{fig:bulk_Ra}(d), and note that bulk value is smaller by nearly an order of magnitude in the active region than the global value: the global thermal dissipation is dominated by contributions from the thermal boundary layers~\citep{Emran:EPJE2012, Bhattacharya:POF2019a}. The global dissipation follows $0.15 Ra^{-0.21}$ scaling, consistent with that of $Nu$ from \eqref{eq:Nu_epst}. The dissipation in the active region, in addition to being weaker, vanishes more rapidly with $Ra$; the best fit yields $\varepsilon_T^{bulk} = 0.04 Ra^{-0.28}$. In the quiet region, $\varepsilon_T^{bulk}$ varies even more rapidly with $Ra$, according to $0.34 Ra^{-0.48}$. These trends confirm our intuition that vanishingly lower thermal dissipation occurs in the quiet region.

\section{Boundary layer properties}
\label{sec:BL}

\subsection{Boundary layer thicknesses}
\label{sec:mean_BL}

\begin{figure}
\captionsetup{width=1\textwidth}
\centering
\includegraphics[width=0.6\textwidth]{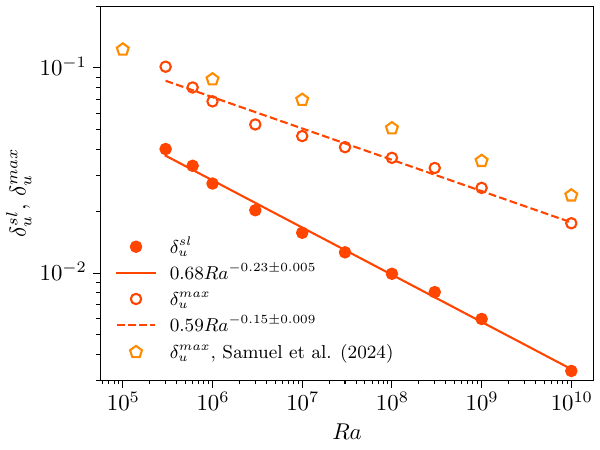}
\caption{Mean thickness of the viscous layer at the horizontal plates, estimated using the slope method (filled red circles) and the maximum method (open red circles), decreases with $Ra$. Orange pentagons represent the $\delta_u^{max}$ in a $\Gamma = 4$ square cuboid from \citet{Samuel:JFM2024}. The maximum method yields a thicker viscous boundary layer as well as a gradual decrease with $Ra$. Red solid and dashed curves are the best fits to $\delta_u^{sl}$ and $\delta_u^{max}$, respectively.}
\label{fig:d_u}
\end{figure}

We compute vertical profiles of the wall-parallel velocity as
\begin{equation}
U = \sqrt{\langle u_x^2 + u_y^2 \rangle_{A,t}} \, ,
\end{equation}
and note that $U$ displays a maximum before decreasing further away into the bulk region. Here $\langle \rangle_{A,t}$ stands for the averaging over the entire horizontal cross-section and time. This maximum horizontal velocity, denoted as $U_{max}$, is used to define the $U_{max}$-based Reynolds number in \S~\ref{sec:global}.
The position of $U_{max}$ from the wall is one estimate for the thickness of the viscous boundary layer (VBL) \citep{Lam:PRE2002,Samuel:JFM2024, Pandey:JFM2024}, and this method of estimating the VBL thickness $\delta_u^{max}$ is referred to as the `maximum method'. We plot the averaged $\delta_u^{max}$ over both the horizontal plates against $Ra$ in figure~\ref{fig:d_u} and find that it follows a power law $\delta_u^{max} = 0.59 Ra^{-0.15}$ over the entire range of $Ra$. On the other hand, the slope thickness, $\delta_u^{sl}$, computed as~\citep{Qiu:PRE1998b}
\begin{equation}
\delta_u^{sl} = \frac{U_{max}}{(\mathrm{d}U/\mathrm{d}z)_{z=0}} \, ,
\end{equation}
and shown in figure~\ref{fig:d_u} as the average from top and bottom plates,
yields a power law $\delta_u^{sl} = 0.68 Ra^{-0.23}$. We thus have the approximate Reynolds number scaling as $\sqrt{Ra}$, the mean VBL thickness varying as $\delta_u^{sl} \sim \sqrt{Re}$. This scaling suggests that the viscous boundary layer retains laminar scaling at least up to $Ra = 10^{10}$~\citep{Blass:JFM2021a, Wagner:JFM2012}.

\begin{figure}
\captionsetup{width=1\textwidth}
\centering
\includegraphics[width=1\textwidth]{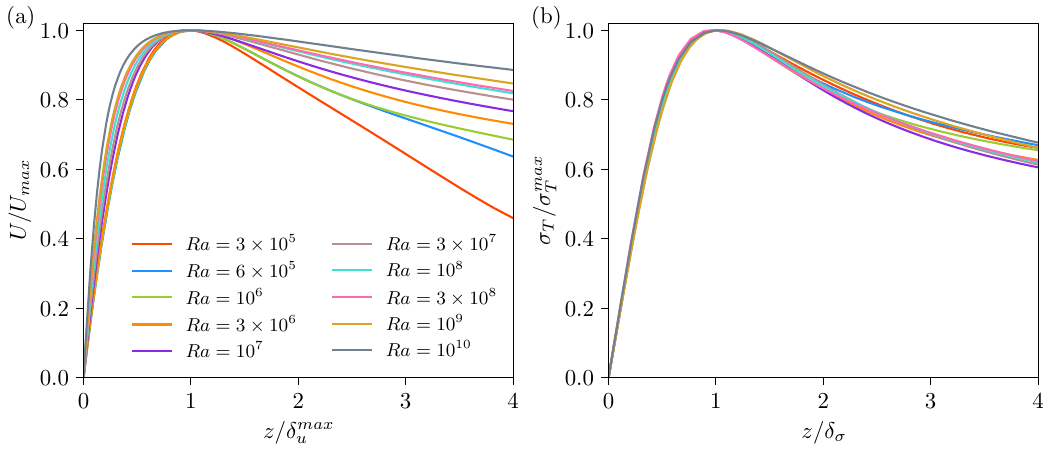}
\caption{(a) Normalised wall-parallel velocity profile $U/U_{max}$ as a function of the normalised distance $z/\delta_u^{max}$ from the plate. The profiles do not exhibit self-similarity within the VBL. (b) Normalised temperature fluctuation profiles as a function of the normalised distance from the plate collapse well in the TBL region. The profiles are additionally averaged over the bottom and top halves of the cell due to the up-down symmetry of Boussinesq convection.}
\label{fig:U_Trms_z_norm}
\end{figure}

In figure~\ref{fig:U_Trms_z_norm}(a), we plot the normalised velocity profiles $U/U_{max}$ as functions of the normalised wall-normal distance $z/\delta_u^{max}$, and note that the profiles do not collapse within the VBL. Thus, a dependence on $Ra$ remains even after the normalisation, which implies that the wall-parallel velocity $U$ does not scale self-similarly in the VBL region.

We now show that, in contrast to velocity profiles, the temperature profiles collapse well within the
TBL. Define the thickness of the thermal boundary layer, $\delta_\sigma$ by the vertical profiles of the RMS temperature fluctuation \citep{Belmonte:PRL1993, Belmonte:PRE1994, Lui:PRE1998}, as
\begin{equation}
\sigma_T = \sqrt{\langle T^2 \rangle_{A,t} - \langle T \rangle_{A,t}^2} \, .
\end{equation}
Figure~\ref{fig:U_Trms_z_norm}(b) shows that the normalised temperature fluctuation $\sigma_T/\sigma_T^{max}$ as a function of the normalised distance $z/\delta_\sigma$ collapse well within the TBL.

\begin{figure}
\captionsetup{width=1\textwidth}
\centering
\includegraphics[width=1\textwidth]{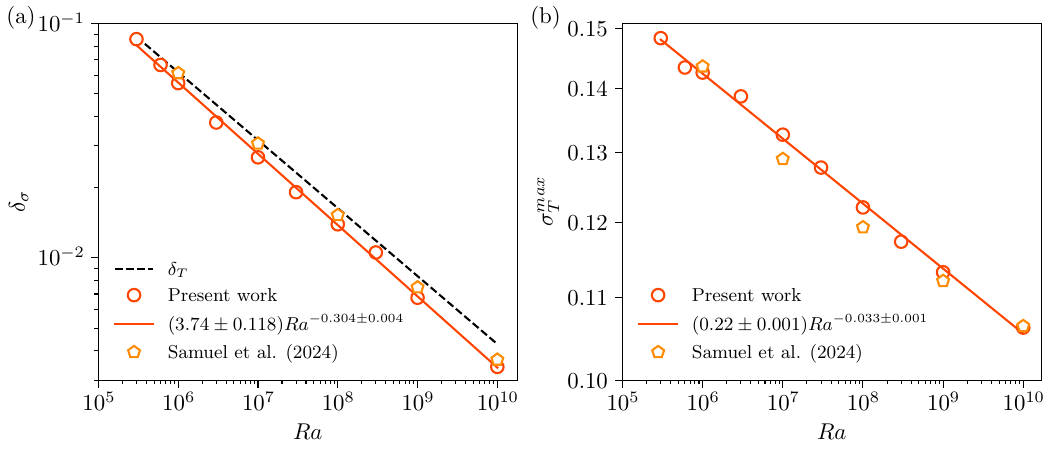}
\caption{(a) Thermal fluctuation thickness $\delta_\sigma$ (symbols) decreases with $Ra$ with the best fit (solid curve) suggesting a $3.74 Ra^{-0.30}$ scaling for the entire range. The $\delta_\sigma$ remains slightly smaller than the slope thickness $\delta_T$ that decreases as $3.4 Ra^{-0.29}$ (dashed line). The present data agrees well with the thermal fluctuation thickness observed for $Pr = 0.7$ in a square cuboid of $\Gamma = 4$~\citep{Samuel:JFM2024}, shown as orange pentagons. (b) Amplitude of the peak temperature fluctuation, $\sigma_T^{max}$, decreases as $Ra^{-0.033}$ in the rectangular cuboid and agrees very well with that observed in the $\Gamma = 4$ square cuboid~\citep{Samuel:JFM2024}. The decreasing trend of $\sigma_T^{max}$ is also similar to that observed in 2D convection for $Pr = 0.021$~\citep{Pandey:JFM2021}.}  
\label{fig:max_Trms}
\end{figure}

We plot the thermal fluctuation thickness $\delta_\sigma$, averaged over the bottom and top plates, as a function of $Ra$ in figure~\ref{fig:max_Trms}(a) and find that it follows $\delta_\sigma \sim 3.74 Ra^{-0.30}$. The mean TBL thickness $\delta_T$ computed using the slope method is nearly the same as the fluctuation thickness. The best fit is $\delta_T = 3.4 Ra^{-0.29}$, shown as a dashed curve in figure~\ref{fig:max_Trms}(a). Recalling the relation $\delta_T = 0.5/Nu$, the variation of $\delta_T$ with $Ra$ is consistent with that of the measured $Nu$. The robustness of $\delta_T$ and $\delta_\sigma$, on the one hand, and differences between $\delta_u^{sl}$ and $\delta_u^{max}$, on the other, show that the TBL can be identified less ambiguously than VBL, probably because the temperature scale is imposed on the system, while the velocity scale results from flow dynamics.

The peak temperature fluctuation with $Ra$ is plotted in figure~\ref{fig:max_Trms}(b); the best fit yields a scaling $\sigma_T^{max} = 0.22 Ra^{-0.033}$. We also include $\sigma_T^{max}$ from the $\Gamma = 4$ square cuboid~\citep{Samuel:JFM2024} in figure~\ref{fig:max_Trms}(b), which follows a scaling $\sigma_T^{max} = 0.22 Ra^{-0.032}$. Thus, the observed scaling of $\sigma_T^{max}$ is quite robust. The decreasing $\sigma_T^{max}$ implies that the plume temperature becomes increasingly similar to that of the ambient fluid as $Ra$ increases~\citep{Pandey:JFM2021}. 

\subsection{Boundary layer evolution on horizontal walls}
\label{sec:str_BL}

The flow near the horizontal plates is inhomogeneous and so we study the profiles extracted at various uniformly spaced locations along the length of the cell at the mid-width $y = L_y/2$. These local profiles are computed as:
\begin{equation}
U^{loc}(x_0) = \sqrt{ \langle  u_x^2 + u_y^2 \rangle_{A_0,t} }
\quad
\mbox{and}
\quad
T^{loc}(x_0) = \langle  T \rangle_{A_0,t} \, .
\end{equation}
Here $A_0$ is a horizontal cross-section of area $0.05H \times 0.05H$ centered at $(x_0,y_0 = L_y/2)$ with $x_0/L_x = \lbrace 2/32, 3/32,..., 30/32 \rbrace$. (We have verified by varying the area of the averaging window between $A_0/4$ and $4 A_0$ that the following results remain robust.) The local profiles were found to be symmetric with respect to the mid-vertical plane $x = L_x/2$, as expected, so we average and discuss them at $x_0 = \lbrace 2 L_x/32, 3 L_x/32,..., L_x/2 \rbrace$. Further, as the ejection region at the bottom plate is located at $x \approx L_x/2$ for some $Ra$, but at $x \approx 0$ or $L_x$ for others, to make the discussion clear, we transform the flow such that $x = L_x/2$ corresponds at all Rayleigh numbers to the ejection and impact regions at bottom and top plates, respectively. The location $x = L_x/4$ continues to be the shear region on both plates (see figure~\ref{fig:sketch}). The horizontal flow is strong in the shear region and the maximum horizontal velocity decreases as one moves away from there~\citep{Qiu:PRE1998a}. The local profiles are asymmetric with respect to the mid-horizontal plane except those in the shear region. 

\begin{figure}
\captionsetup{width=1\textwidth}
\centering
\includegraphics[width=1\textwidth]{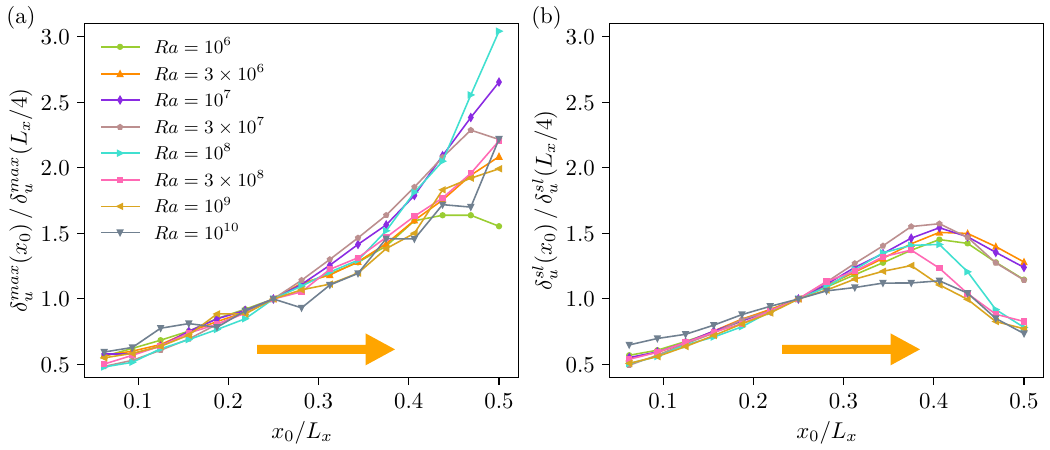} 
\caption{Structure of the viscous boundary layer at the bottom plate. (a) Local thickness estimated using the maximum method normalised with that in the shear region increases in the direction of the wind (indicated by orange arrows). (b) Thickness estimated using the slope method exhibits an increasing trend up to $x_0 \approx 0.4 L_x$, beyond which it decreases.}
\label{fig:d_u_loc}
\end{figure}

\begin{figure}
\captionsetup{width=1\textwidth}
\centering
\includegraphics[width=1\textwidth]{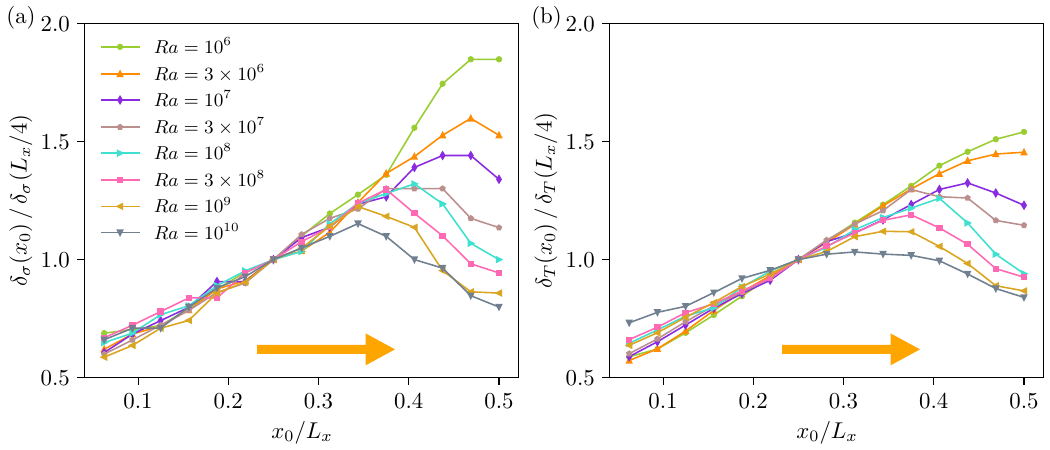} 
\caption{Structure of the thermal boundary layer at the bottom plate. Both the local fluctuation thickness (a) as well as the slope thickness (b) normalised with that in the shear region increases in the direction of the wind (indicated by orange arrows). However, the increasing trend breaks down at larger Rayleigh numbers beyond $x_0 \approx 0.4 L_x$.}
\label{fig:d_T_loc}
\end{figure}

We show in figures~\ref{fig:d_u_loc} and ~\ref{fig:d_T_loc} the local boundary layer thicknesses computed from these profiles, normalised by that in the shear region, for normalised distances $x_0/L_x$. To avoid clutter, we show data only at the bottom plate. Figure~\ref{fig:d_u_loc} shows the variation of the VBL thickness estimated using the maximum method in panel (a), and that using the slope method in panel (b). Both methods yield local VBL thicknesses that increase, up to $x_0/L_x \approx 0.4$ in the wind direction, indicated using orange arrows. See also \citet{Qiu:PRE1998b} and \citet{Wagner:JFM2012} for similar observations. Figure~\ref{fig:d_u_loc}(b) shows that $\delta_u^{sl}(x_0)$ starts to decrease as the ejection region is approached, but not the $\delta_u^{max}(x_0)$.As the ejection region in the {present configuration} is far from the sidewalls, it is possible that the decline of the local VBL thickness is an artifact of the slope method. Clusters of thermal plumes with locally smaller spacing detach at the edge of the ejection region rather than from its center, where a stagnation region forms due to colliding mean flows. This behaviour can be probed only by the slope method that builds on vertical temperature derivatives at the wall. With increasing $Ra$, larger number of finer-scale plumes aggregates, causing broader and less coherent clustering regions, thus decreasing thickness ratios with increasing $Ra$.

As shown in figure~\ref{fig:d_T_loc}, TBL thicknesses from both methods increase to $L_x/2$ up to $Ra = 3 \times 10^6$ but decreases with $x_0$ in the region $x_0 \geq 0.4 L_x$ for higher $Ra$. It is also clear that the variation of the TBL thickness over the plate becomes milder as $Ra$ increases, consistent with earlier observations~\citep{Lui:PRE1998, Pandey:JFM2021}. Figure~\ref{fig:d_u_loc} further suggests that the VBL thickness varies over the plate more strongly than the TBL thickness. 

\begin{figure}
\captionsetup{width=1\textwidth}
\centering
\includegraphics[width=1\textwidth]{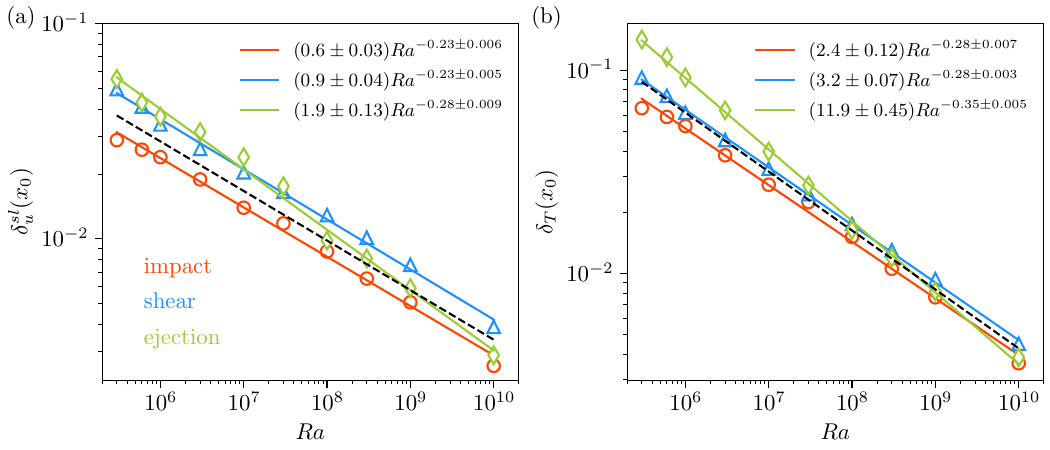}
\caption{Scaling of (a) local VBL thickness and (b) TBL thickness as functions of $Ra$. Red circles are data for the impact region ($x_0 = L_x/2$ at the top plate), blue triangles are for shear region ($x_0 = L_x/4$ and $3 L_x/4$), and green diamonds are for ejection region ($x_0 = L_x/2$ at the bottom plate). Best fits reveal that the exponents are the same for impact and shear regions and smaller than that for ejection region. Black dashed curve in both (a) and (b) stand for the variation of the mean boundary layer thickness.}
\label{fig:d_u_loc_ra}
\end{figure}

In figure~\ref{fig:d_u_loc_ra}(a), we plot the VBL thicknesses $\delta_u^{sl}(x_0)$ in the impact, shear, and ejection regions as functions of $Ra$, and observe that, as expected, $\delta_u^{sl}(x_0)$ decrease with $Ra$ everywhere. The best fit results in a scaling $\delta_u^{sl}(x_0) = 0.6 Ra^{-0.23}$ in the impact region ($x \approx L_x/2$ at the top plate). The thicknesses in the shear region (blue triangles) are slightly larger but the scaling exponent is nearly the same. In this region, VBL thickness scales as $\delta_u^{sl}(x_0) = 0.9 Ra^{-0.23}$. Note that the thickness in the shear region corresponds to the average thickness at $x_0 = L_x/4$ and $x_0 = 3L_x/4$ both at the bottom and top plates. For comparison, we also show the scaling of the mean VBL thickness $\delta_u^{sl} = 0.68 Ra^{-0.23}$ as a black dashed curve in figure~\ref{fig:d_u_loc_ra}(a). The VBL thickness in the impact (shear) region remains consistently smaller (larger) than the mean thickness at the entire horizontal plate. The thickness in the ejection region (at $x \approx L_x/2$ at the bottom plate) declines more rapidly with $Ra$, with the best fit yielding $1.9 Ra^{-0.28}$.

Scalings of the local TBL thicknesses are shown in figure~\ref{fig:d_u_loc_ra}(b). As for VBL thicknesses, scalings in impact and shear regions are similar and differ from that in ejection region. The best fits result in the following scalings: $\delta_T(x_0) = 2.4 Ra^{-0.28}$ in impact region, $\delta_T(x_0) = 3.2 Ra^{-0.28}$ in shear region, and $\delta_T(x_0) = 11.8 Ra^{-0.35}$ in ejection region. Local thermal fluctuation thicknesses, $\delta_\sigma(x_0)$, scale very similarly to $\delta_T(x_0)$ in these regions and thus also exhibit a strong difference between the scaling exponents in the ejection and shear regions. The scaling of TBL thickness in shear region is nearly the same as that for the mean thickness over the entire plate, which is shown as a black dashed curve in figure~\ref{fig:d_u_loc_ra}(b). A rapid decrease of TBL thickness in ejection region, compared to those in shear and impact regions, is similar to that observed in low-$Pr$ convection, both in 2D~\citep{Pandey:JFM2021} and 3D~\citep{Kim:JFM2024}. As the heat transport is inversely proportional to the TBL thickness, this difference also implies that the local heat transport in the ejection region grows more rapidly with $Ra$ than that in the shear region (\citet{Zhu:PRL2018}) for $Pr = 1$ in {3D} RBC.    

\section{Conclusions} 
\label{sec:concl}

In convection studies, it is customary to separate the flow into a boundary layer region and the bulk. There is, in general, no ambiguity to the notion when the aspect ratio is very large, which was the initial intent. When applying the concepts to small aspect ratios, the division is not conceptually clear-cut because the flow near the boundary is strongly influenced by the vertical walls of the container. We have studied convection where the vertical walls are far enough away from each other so as to reduce their influence on the boundary layer. In this configuration, it is possible to study regions dominated by a strong activity of thermal plumes, quiet regions, as well as regions of shear and impact. The configuration we have chosen is a rectangular cuboid of length $2.4H$ and width $0.8H$, and DNS were performed for $Pr = 1$ and $3 \times 10^5 \leq Ra \leq 10^{10}$. For these conditions, the walls do not strongly influence the plume dynamics, so that the flow characteristics can be explored in plume-abundant and plume-depleted regions, as well as regions of impact and shear.

What we learn is that, while the thermal boundary layer can be unambiguously identified and is self-similar {with respect to $Ra$}, the velocity boundary layer is not; its characteristic thickness depends on the definition. We observed, further, that the horizontal velocity changes gradually or rapidly from the wall depending on whether one considers the ejection, impact or shear region.

We explored the properties of velocity and temperature fluctuations and dissipation rates in active and quiet regions, located deeper in the bulk of the convective layer. The velocity fluctuations exhibit nearly similar characteristics in all these regions, this property being true for the kinetic energy dissipation rate as well. On the other hand, temperature fluctuations, as well as the thermal dissipation rate do depend on the region of the flow where they are measured. We find that the scalings of all thermal quantities in the active region are shallower (i.e. the corresponding scaling exponents have a smaller magnitude) than those in the quiet region.

We further found that both the viscous and thermal boundary layers become thicker in the direction of the mean wind, though their structure in the ejection region is not monotonic. The behaviour in the ejection region was observed to depend on the definition of the boundary layer thickness in the case of the viscous layer, whereas only on the Rayleigh number in case of the thermal layer. We also studied the scalings of the local boundary layer thicknesses with respect to $Ra$ and observed that both VBL and TBL become thinner more rapidly in the ejection region than those in the shear and impact regions. These results form the bulk of the paper.

Our study also led us to two general qualitative lessons. First, the boundary layers and the bulk are intimately intertwined for the parameter ranges explored here, even though both boundary layers get increasingly thinner with increasing $Ra$. Thus, theories that consider the bulk or the boundary layers as {more} important than the other cannot be distinguished for their accuracy.  Second, despite these differences of detail, the global heat transport laws are practically the same as those in other configurations of low to moderate aspect ratios. In other words, focusing merely on the Nusselt number characteristics does not help us understand the nature of the flow adequately.



\backsection[Acknowledgements]{This research was carried out on the High Performance Computing resources at New York University Abu Dhabi. The authors also gratefully acknowledge {\sc Shaheen} II of KAUST, Saudi Arabia (under project numbers k1491 and k1624) for providing computational resources. }

\backsection[Funding]{This material is based upon work supported by Tamkeen under the NYU Abu Dhabi Research Institute grant G1502, and by the KAUST Office of Sponsored Research under Award URF/1/4342-01. A.P. also acknowledges financial support from ANRF (formerly SERB), India under the grant SRG/2023/001746. NYU supports the research of K.R.S.}

\backsection[Declaration of Interests]{The authors report no conflict of interest.}

\backsection[Data availability statement]{The data that support the findings of this study are available from the corresponding author upon reasonable request.}

\backsection[Author ORCIDs]{\\
A. Pandey, \href{https://orcid.org/0000-0001-8232-6626}{https://orcid.org/0000-0001-8232-6626};\\
J. Schumacher, \href{https://orcid.org/0000-0002-1359-4536}{https://orcid.org/0000-0002-1359-4536}; \\
M. Parsani, \href{https://orcid.org/0000-0001-7300-1280}{https://orcid.org/0000-0001-7300-1280}; \\
K. R. Sreenivasan, \href{https://orcid.org/0000-0002-3943-6827}{https://orcid.org/0000-0002-3943-6827}.
}


\end{document}